%% file: main.tex
\documentclass{article}
\usepackage{iclr2027_conference,times}

\usepackage{amsmath,amssymb,mathtools,bm}
\usepackage{booktabs}
\usepackage{array}
\usepackage{graphicx}
\usepackage{xcolor}
\usepackage{framed}
\usepackage{fvextra}
\usepackage{microtype}
\usepackage{multirow}
\usepackage[section]{placeins}
\usepackage{tabularx}
\usepackage{xspace}
\usepackage{url}
\usepackage{hyperref}

\let\cite\citep

\input{paper_commands.tex}

\title{\inspire: Benchmarking Scientific Literature Search for Open Research Problems}

\author{%
Jianrong Ding$^{1}$\thanks{Work done during Jianrong Ding's internship at Microsoft Research Asia.}\hphantom{$^{*}$},
Zhengyan Shi$^{1}$,
Jianyuan Zhong$^{1}$,
Kai Qiu$^{2}$,
\\
\bfseries Qi Dai$^{2}$,
Yifan Yang$^{2}$,
Chong Luo$^{2}$,
Qiang Xu$^{1}$\thanks{Corresponding author.}
\\
$^{1}$Department of Computer Science and Engineering, The Chinese University of Hong Kong
\\
$^{2}$Microsoft Research Asia
\\
\texttt{\{jrding25,zyshi25,jyzhong,qxu\}@cse.cuhk.edu.hk}
\\
\texttt{\{kaqiu,qid,yifanyang,cluo\}@microsoft.com}
}

\iclrfinalcopy 

\begin{document}

\maketitle

\input{sections/abstract}
\input{sections/introduction}

\input{sections/coupled_search}
\input{sections/inspire_benchmark}
\input{sections/experiment}

\input{sections/results}
\input{sections/related_work}
\input{sections/limitations}
\input{sections/conclusion}

\input{sections/ai_disclosure}

\bibliography{references}
\bibliographystyle{iclr2027_conference}

\clearpage
\appendix

\input{sections/dataset_details}

\input{sections/experiment_details}
\input{sections/hindsight_sft}
\input{sections/oracle_search_progress}
\input{sections/llm_prompt_templates}

\end{document}

%% file: paper_commands.tex
\newcommand{\inspire}{\textsc{Inspire}\xspace}
\newcommand{\timemachine}{\textsc{TimeMachine}\xspace}
\newcommand{\sftmodel}{Qwen3.6-35B-A3B\xspace}
\newcommand{\ndcg}{\operatorname{nDCG}}

\newcommand{\anyhit}{\ensuremath{\operatorname{AnyHit@10}}\xspace}

\newcommand{\cmark}{\ensuremath{\bm{\checkmark}}}
\newcommand{\xmark}{\ensuremath{\bm{\times}}}

\definecolor{InspireInk}{HTML}{25313C}
\definecolor{InspireMuted}{HTML}{66737F}
\definecolor{InspireNavy}{HTML}{315B7D}
\definecolor{InspireBlue}{HTML}{5B8DB8}
\definecolor{InspireTeal}{HTML}{4C9A8A}
\definecolor{InspireOrange}{HTML}{D8921B}
\definecolor{InspireRed}{HTML}{B85C5C}
\definecolor{InspireLightBlue}{HTML}{EAF2F8}
\definecolor{InspireLightTeal}{HTML}{EAF5F2}
\definecolor{InspireLightOrange}{HTML}{FBF3E3}
\definecolor{InspireLightRed}{HTML}{FDEEEE}
\definecolor{InspirePanelGray}{HTML}{F3F5F6}
\definecolor{InspireBorderGray}{HTML}{CAD1D6}

\newcommand{\figureorplaceholder}[3]{%
  \IfFileExists{#1}{%
    \includegraphics[width=#2]{#1}%
  }{%
    \fbox{\parbox[c][#3][c]{0.92\linewidth}{\centering\small
      Figure asset placeholder\\[0.4em]
      \texttt{\detokenize{#1}}}}%
  }%
}

%% file: sections/abstract.tex
\begin{abstract}

Scientific literature search often begins with an open research problem rather than a known target paper or a fixed candidate set. We introduce \inspire, a benchmark for evaluating agents that search prior literature to make progress on solution-redacted research problems. Each instance pairs a research brief with a target-specific cutoff three months before a later paper and evaluates ranked outputs against graded cited antecedents from that paper's realized research lineage. Search proceeds over an open corpus, while the identity of the target paper and membership of its cited antecedents remain hidden from the agent. Beyond end-to-end retrieval quality, \inspire uses logged search trajectories to distinguish three coupled stages: \emph{resource exposure}, whether useful antecedents are surfaced during search; \emph{selection}, whether exposed antecedents are retained; and \emph{ranking}, how effectively retained papers are ordered. Across 476 computer-science targets under a shared search interface and budget, the strongest evaluated agent achieves 0.284 nDCG@10. Results show that current agents more readily recover an isolated antecedent than assemble a broader portfolio of relevant prior work. The stagewise analysis identifies resource exposure as the largest observed bottleneck, with further losses in selection and ranking. We additionally construct replay-valid hindsight demonstrations and show that they improve held-out search without changing test-time information, establishing that the benchmark provides an actionable learning signal. \inspire therefore enables both end-to-end comparison and stage-resolved diagnosis in a setting where the agent must construct its own working criterion of relevance.

\end{abstract}

%% file: sections/introduction.tex
\section{Introduction}
\label{sec:introduction}

Search-augmented language models have progressed from retrieval-augmented generation and browser-assisted question answering to agents that interleave reasoning, tool use, evidence inspection, and synthesis \cite{lewis2020rag,nakano2021webgpt,yao2023react,schick2023toolformer,liu2024agentbench,wei2025browsecomp,du2025deepresearchbench}.  Reliable scientific search, however, is not a single retrieval decision.  It comprises three stages: \emph{resource exposure}, in which the agent searches broadly enough to surface useful candidate papers; \emph{selection}, in which it retains the most relevant papers from those exposed; and \emph{ranking}, in which it orders the retained evidence into a compact final response.  These stages are diagnostically distinct but operationally coupled.  When search begins from an unresolved problem, the eventual solution is hidden, useful mechanisms may not share the problem's vocabulary, and the agent's provisional relevance criterion governs both which evidence becomes visible and which evidence is retained.  Retrieved evidence can in turn revise that criterion and redirect subsequent search.  This evolving information need is central to anomalous-state, berrypicking, and exploratory-search accounts \cite{belkin1982ask,bates1989berrypicking,marchionini2006exploratory}, and to the view that relevance is contextual \cite{saracevic1975relevance}.

Existing benchmarks evaluate important subsets of this process, but often stabilize what counts as correct before or during search: BrowseComp supplies clues, AutoResearchBench explicit acceptance conditions, PaperPilot an anchor and optional user clarification, and ResearchBench a curated candidate set \cite{wei2025browsecomp,autoresearchbench2026,li2026paperpilot,liu2025researchbench}.  MIR and HyBIRD are closer problem-conditioned inspiration retrievers, while PreScience evaluates future prior-work choices using author and publication histories \cite{garikaparthi2025mir,yang2026hybird,ajith2026prescience}.  We classify a task that specifies identifying information and asks for the matching paper as an evaluation of resource exposure: once the paper is surfaced, the supplied conditions determine whether it should be returned.  By contrast, selection requires the benchmark to evaluate which already observed candidates the model retains or rejects.  Existing protocols may evaluate exposure, selection, or ranking individually, but do not measure all three on a common trajectory; consequently, their end-to-end scores cannot determine where relevant evidence was lost.

\input{tables/benchmark_comparison}
\begin{figure*}[!t]
  \centering
  \figureorplaceholder{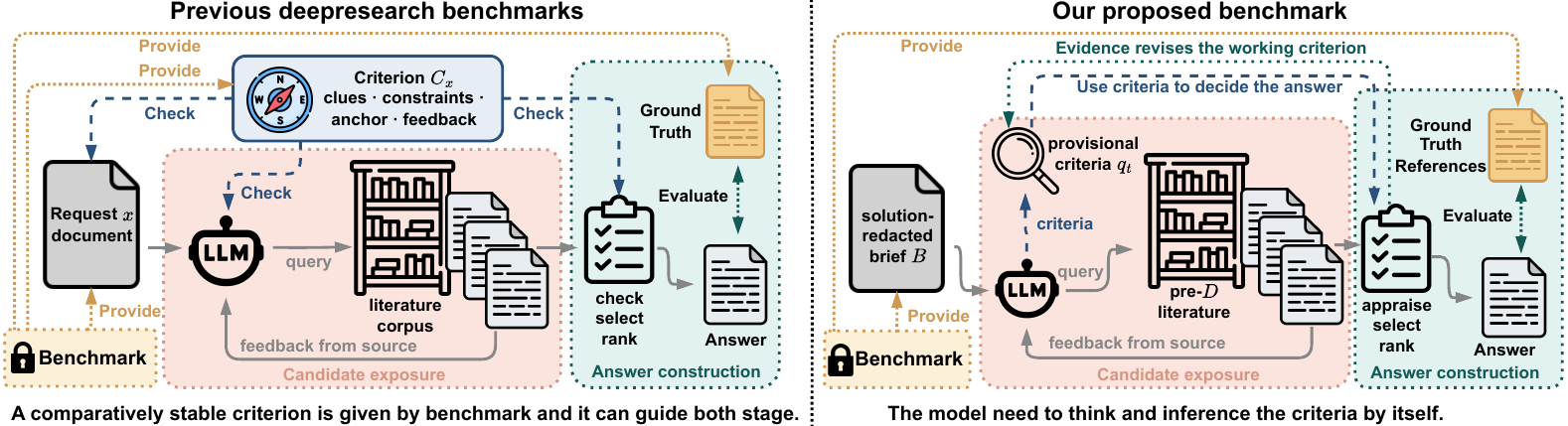}{0.99\textwidth}{2.35in}
  \caption{\textbf{Source of the working relevance criterion.}  Left: prior tasks stabilize relevance through clues, constraints, anchors, feedback, or candidate pools.  Right: \inspire supplies only a solution-redacted problem, so the agent must form provisional criteria that guide resource exposure, selection, and ranking.  Retrieved evidence can revise those criteria and redirect subsequent search.  The schematic contrasts information supplied during interaction; task difficulty can vary independently.}
  \label{fig:compare}
\end{figure*}

Table~\ref{tab:benchmark-comparison} compares benchmark characteristics under these definitions.  \inspire evaluates all three stages of scientific search within a single open-corpus trajectory.  Given only a solution-redacted problem brief, an agent searches literature published before a target-specific cutoff and submits up to ten ranked papers, without access to the hidden target, its bibliography, or relevance labels.  After the episode, graded cited antecedents from the target's realized lineage provide three nested measurements: \emph{exposed gain} evaluates resource exposure, \emph{selected gain} evaluates which exposed papers are retained, and final nDCG@10 evaluates their ranking.  The gaps between these measurements reveal whether useful evidence was never surfaced, was exposed but omitted, or was retained but poorly ordered.

Under a shared interface and budget, the strongest model reaches 0.284 nDCG@10 and recovers at least one antecedent on 72.3\% of targets, yet recovers three on only 32.1\% of eligible targets.  Across models, the zero-exposure rate ranges from 17.4\% to 56.1\%, and the largest observed loss occurs before selection.  A matched hindsight-SFT intervention raises held-out nDCG@10 from 0.082 to 0.166 under unchanged test-time information, showing that \inspire supplies an actionable learning signal.  Figure~\ref{fig:compare} contrasts this information structure with supplied-criterion tasks.
Our contributions are:
\begin{itemize}
    \item We formulate scientific search as three coupled stages---resource exposure, selection, and ranking---and show how a provisional relevance criterion can shape all three.
    \item We introduce \inspire, a temporally isolated, solution-redacted benchmark that evaluates each stage from logged trajectories using hindsight-revealed graded antecedents while strictly separating policy and evaluator information.
    \item We provide a controlled multi-model comparison that localizes current failures across the three-stage funnel, and a matched Base--SFT intervention showing that \inspire's training signal produces measurable improvement on held-out targets.
\end{itemize}

%% file: tables/benchmark_comparison.tex
\begin{table*}[!t]
  \caption{\textbf{Capability coverage of selected scientific-search benchmarks.}
  A \textcolor{InspireTeal}{\cmark} denotes coverage and a \textcolor{InspireRed}{\xmark} denotes no coverage under our operational definitions.  Resource exposure evaluates whether a system can surface the needed literature from a corpus.  Selection requires evaluating which already exposed papers are retained or omitted, and ranking evaluates their submitted order. HyBIRD uses MIR~\cite{garikaparthi2025mir}.  Only \inspire covers all five characteristics.}
  \label{tab:benchmark-comparison}
  \centering
  \small
  \renewcommand{\arraystretch}{1.12}
  \setlength{\tabcolsep}{3.8pt}
  \resizebox{\textwidth}{!}{%
  \begin{tabular}{lccccc}
    \toprule
      & \multicolumn{2}{c}{Task setting}
      & \multicolumn{3}{c}{Research capability evaluated} \\
    \cmidrule(lr){2-3}\cmidrule(lr){4-6}
    Benchmark
      & \shortstack{Interactive\\open-corpus search}
      & \shortstack{No supplied\\relevance cue}
      & \shortstack{Resource\\exposure}
      & Selection
      & Ranking \\
    \midrule
    AutoResearchBench~\cite{autoresearchbench2026}
      & \textcolor{InspireTeal}{\cmark} & \textcolor{InspireRed}{\xmark}
      & \textcolor{InspireTeal}{\cmark} & \textcolor{InspireRed}{\xmark}
      & \textcolor{InspireRed}{\xmark} \\
    PaperPilot~\cite{li2026paperpilot}
      & \textcolor{InspireTeal}{\cmark} & \textcolor{InspireRed}{\xmark}
      & \textcolor{InspireTeal}{\cmark} & \textcolor{InspireRed}{\xmark}
      & \textcolor{InspireTeal}{\cmark} \\
    ResearchBench~\cite{liu2025researchbench}
      & \textcolor{InspireRed}{\xmark} & \textcolor{InspireRed}{\xmark}
      & \textcolor{InspireRed}{\xmark} & \textcolor{InspireTeal}{\cmark}
      & \textcolor{InspireRed}{\xmark} \\
    HyBIRD~\cite{yang2026hybird}
      & \textcolor{InspireRed}{\xmark} & \textcolor{InspireTeal}{\cmark}
      & \textcolor{InspireTeal}{\cmark} & \textcolor{InspireRed}{\xmark}
      & \textcolor{InspireTeal}{\cmark} \\
    PreScience~\cite{ajith2026prescience}
      & \textcolor{InspireRed}{\xmark} & \textcolor{InspireRed}{\xmark}
      & \textcolor{InspireTeal}{\cmark} & \textcolor{InspireRed}{\xmark}
      & \textcolor{InspireTeal}{\cmark} \\
    \midrule
    \textbf{\inspire (ours)}
      & \textcolor{InspireTeal}{\cmark} & \textcolor{InspireTeal}{\cmark}
      & \textcolor{InspireTeal}{\cmark} & \textcolor{InspireTeal}{\cmark}
      & \textcolor{InspireTeal}{\cmark} \\
    \bottomrule
  \end{tabular}}
\end{table*}

%% file: sections/coupled_search.tex
\section{A Coupled View of Agentic Search}
\label{sec:coupled-search}

\subsection{Resource Exposure, Selection, and Ranking}

\paragraph{Notation.}
Let $B$ denote a problem brief, $\mathcal C$ the searchable corpus, $E\subseteq\mathcal C$ the papers exposed during search, $S=(s_1,\ldots,s_m)$ the submitted ranking, $\bar S=\{s_1,\ldots,s_m\}$ its unordered selected set, and $Y\subseteq\mathcal C$ the hidden graded label set.  Omitting target subscripts, success requires
\begin{equation}
  Y\cap E\neq\varnothing
  \quad\text{and}\quad
  Y\cap \bar S\neq\varnothing,
  \label{eq:two-requirements}
\end{equation}
with stronger portfolio tasks requiring several members of $Y$.  The three stages ask progressively stricter questions: did a graded paper enter $E$ (resource exposure), did it survive in $\bar S$ (selection), and how effectively was $\bar S$ ordered in $S$ (ranking)?  All three are observable from returned and submitted identifiers.  Zero exposure establishes a hard ceiling for any answer constructed from the observed trajectory, whereas positive exposure is only a necessary condition: the trajectory may contain a low-gain antecedent, omit it from the final response, or place it at an ineffective rank.  This nested structure enables diagnostic evaluation even though the stages interact.
These stages are coupled rather than a retrieve-then-rank pipeline: an evolving information need shapes queries and appraisal, while retrieved evidence changes that need \cite{belkin1982ask,bates1989berrypicking,marchionini2006exploratory}.  Let $H_t$ be the history, $z$ a relevance hypothesis (such as a candidate mechanism family), and $q_t(z)$ the agent's distribution over hypotheses.  The same $q_t$ can govern search, appraisal of paper $d$, and subsequent belief updates:
\begin{align}
  a_t^{\mathrm{search}} &\sim
    \pi_{\mathrm{search}}(\,\cdot\mid B,H_t,q_t), \\
  u_t(d) &= \mathbb E_{z\sim q_t}[r(d\mid B,H_t,z)], \\
  q_{t+1} &= \mathcal U(q_t;B,H_t,O_t),
  \label{eq:coupled-search}
\end{align}
where $a_t^{\mathrm{search}}$, $\pi_{\mathrm{search}}$, $u_t$, $r$, $O_t$, and $\mathcal U$ are the next action, search policy, utility, relevance score, observation, and update rule.  We use this analysis model to express how one provisional criterion affects both visibility and submission.  A retrieved paper can shift mass toward a new mechanism family, changing the next query and the appraisal of papers already seen.  The logged stages therefore support diagnostic accounting while retaining their dependence on a shared, evolving criterion.

\subsection{Supplied and Endogenous Relevance Criteria}

Many tasks supply a comparatively stable account of what qualifies through clues, conditions, an anchor, user feedback, or candidate construction; BrowseComp, AutoResearchBench, PaperPilot, and ResearchBench instantiate these sources differently \cite{wei2025browsecomp,autoresearchbench2026,li2026paperpilot,liu2025researchbench}.  Checking such a criterion may still require full-text reading, multi-hop search, or a learned judge; the distinction concerns information structure rather than intrinsic difficulty.  In \inspire, by contrast, the visible problem may support distinct solution lineages $z_1$ and $z_2$ with different antecedent sets $Y(z_1)\ne Y(z_2)$.  The evaluator uses the later realized $Y(z^\star)$, but $z^\star$ is hidden during search and the pre-cutoff record provides no membership certificate.  We use \emph{endogenous} in this operational sense: the agent forms the working criterion that drives its trajectory, and hindsight supplies an exact graded score after the episode.
MIR and HyBIRD are related problem-conditioned inspiration retrievers \cite{garikaparthi2025mir,yang2026hybird}, and PreScience also predicts later prior-work choices without an in-episode certificate \cite{ajith2026prescience}.  \inspire's distinction is the conjunction of solution redaction, interactive date-gated open-corpus search, no anchor or clarification, hindsight evaluation, and logged stage measurements.  The resulting funnel locates observed losses among exposure, selection, and ranking; causal diagnosis and counterfactuals under complete exposure require intervention.

%% file: sections/inspire_benchmark.tex
\section{\inspire Benchmark}
\label{sec:inspire}

\subsection{Task}
\label{sec:task}

For target paper $P_i$ first posted at $t_{P_i}$, instance $i$ is
\begin{equation}
  x_i=(B_i,D_i,\mathcal C_{<D_i},Y_i,g_i),
  \qquad D_i=t_{P_i}-\Delta,
  \label{eq:instance}
\end{equation}
where $\Delta=3$ months, $B_i$ is the solution-redacted brief, $D_i$ the cutoff, $\mathcal C_{<D_i}$ the searchable literature, $Y_i$ the hidden graded cited antecedents, and $g_i$ their gain function, with zero gain outside $Y_i$.  A policy sees only $B_i$ and its tool history; it never observes $P_i$, labels, future citation contexts, or evaluation feedback.  The target-specific cutoff prevents the agent from using the target itself, concurrent work, or later public evidence, while retaining the open-corpus character of the search episode.

The ranked submission is $S_i=(s_{i1},\ldots,s_{im_i})$, with $m_i\leq K=10$.  Version suffixes are normalized, duplicates retain their first position, and malformed, post-cutoff, or unlabeled IDs receive zero.  The primary score is graded nDCG@10:
\begin{equation}
  \ndcg_{K,i}(S_i)=
  \frac{\sum_{j=1}^{\min(K,m_i)}g_i(s_{ij})/\log_2(j+1)}
       {\sum_{j=1}^{\min(K,|Y_i|)}g_i(y_{ij}^\star)/\log_2(j+1)},
  \label{eq:ndcg}
\end{equation}
where $(y_{i1}^\star,\ldots)$ orders $Y_i$ by decreasing gain \cite{jarvelin2002cumulated}.  The target-specific denominator compares a submission with the ideal ordering of the same hidden label set, so scores remain normalized despite variation in label count and gain.  Recall, reciprocal rank, and one/three/five-hit rates complement this view; failures and non-submissions remain zero.  Because citations have heterogeneous functions and omit influences \cite{teufel2006automatic,valenzuela2015meaningful,macroberts1989problems,bornmann2008citation}, the construct is recovery of graded antecedents from one realized lineage; scientific usefulness beyond that lineage and causal credit lie outside the score.

\subsection{Construction}
\label{sec:construction}

\begin{figure*}[!t]
  \centering
  \figureorplaceholder{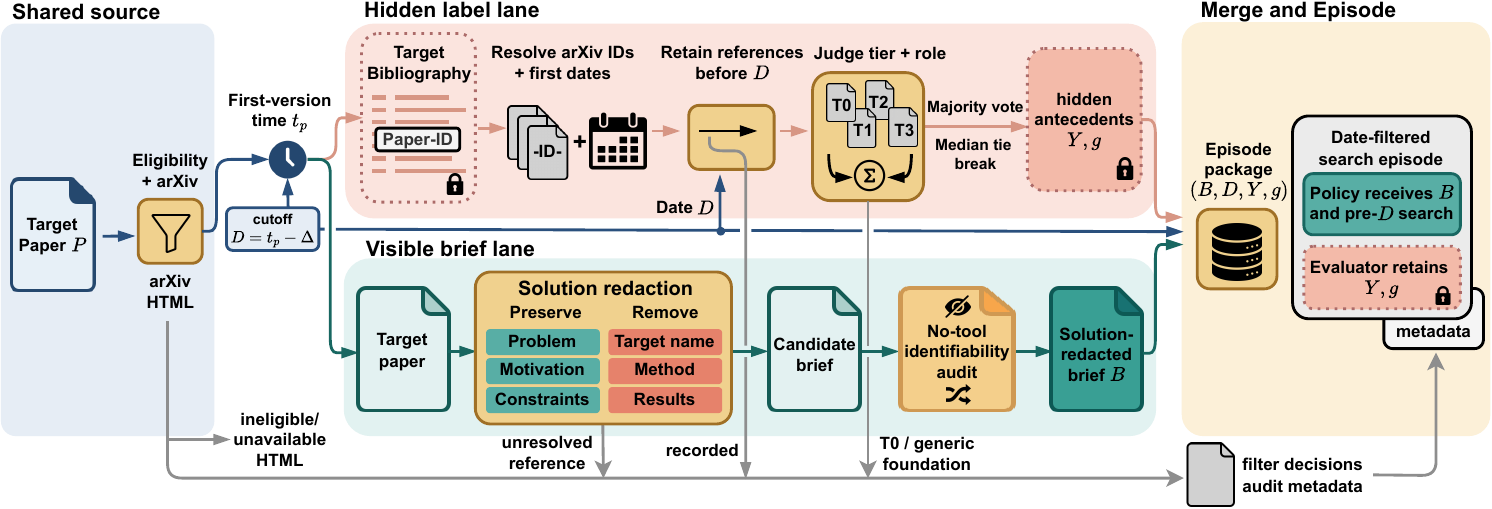}{0.98\textwidth}{2.35in}
  \caption{\textbf{Backward construction and information boundary.}  Starting from a later target, the builder applies the three-month cutoff, reconstructs and screens a solution-redacted brief, and grades cited pre-cutoff antecedents.  During evaluation, the policy receives only the brief and date-filtered search results; the target, its bibliography, and all grades remain evaluator-only.  This separation permits hindsight scoring without exposing the realized solution lineage.}
  \label{fig:construction}
\end{figure*}

Construction proceeds backward from recent, influential computer-science target papers.  We discover targets through topic-stratified Semantic Scholar searches, deduplicate arXiv identifiers, and use arXiv metadata as the authority for title, abstract, category, and first-posting date; surveys, reviews, papers without parseable HTML, and targets with fewer than two graded pre-cutoff antecedents are excluded.  For each retained target $P_i$, we fix $D_i=t_{P_i}-3$ months, normalize arXiv identifiers in its bibliography, and exclude references first posted on or after the cutoff.  Role-sensitive judges then grade each eligible reference using the target abstract, candidate metadata, and citation context when available, producing consensus tier, role, rationale, and gain; non-antecedents are removed and the graded set is capped at ten.  In parallel, a generator converts the target title and abstract into a problem brief that preserves the research motivation and constraints while removing the target identity, realized mechanism, and target-specific results.  A no-tool identity audit and a lexical-overlap check define reproducible clean and flagged diagnostic partitions.  The released instance exposes only the brief, cutoff, and date-filtered search interface to the policy, while the target, bibliography, grades, and citation evidence remain evaluator-only; the manifest records split membership, construction settings, instance hashes, and judge versions.  This design yields scalable silver labels for a realized citation lineage without leaking that lineage into the search episode.  Appendix~\ref{app:data} provides full construction and sensitivity details, and Figure~\ref{fig:construction} illustrates the information boundary.


\subsection{Stage-Wise Metrics}
\label{sec:funnel-protocol}

End-to-end scores cannot distinguish a relevant paper that was never observed from one that was exposed and then omitted or poorly ranked.  We therefore introduce a nested evaluation of resource exposure, selection, and ranking.  For each of $N$ targets, let $E_i$ be the set of unique paper identifiers returned during the trajectory, and let $\bar S_i=\{s_{ij}\}_{j=1}^{m_i}$ be the unordered support of the ranked submission $S_i$.  The decomposition uses only papers observed in the logged trajectory: it neither retrieves new candidates nor changes the policy's information.

For any set $A\subseteq\mathcal C_{<D_i}$, define its normalized available gain by optimally ordering its labeled members:
\begin{equation}
  \phi_i(A)=\ndcg_{K,i}\!\left(\operatorname{sort}_{g_i}(A\cap Y_i)\right)\in[0,1].
\label{eq:ranking-potential}
\end{equation}
Thus, $\phi_i(E_i)$ is the best score supported by the observed exposure set, rather than the outcome of a counterfactual search policy.  We aggregate performance after each stage as
\begin{equation}
\begin{aligned}
G_{\mathrm{exposed}}&=\frac{1}{N}\sum_i\phi_i(E_i), &
G_{\mathrm{selected}}&=\frac{1}{N}\sum_i\phi_i(\bar S_i\cap E_i), &
G_{\mathrm{final}}&=\frac{1}{N}\sum_i\ndcg_{K,i}(S_i).
\end{aligned}
\label{eq:funnel-stages}
\end{equation}
$G_{\mathrm{exposed}}$ measures the ranking potential made available through resource exposure; $G_{\mathrm{selected}}$ measures the potential retained after selection, independent of submitted order; and $G_{\mathrm{final}}$ measures the realized quality after ranking.  Because submissions must be drawn from exposed papers,
\begin{equation}
G_{\mathrm{final}}\leq G_{\mathrm{selected}}\leq G_{\mathrm{exposed}}\leq 1.
\label{eq:funnel-order}
\end{equation}
The corresponding absolute losses are $1-G_{\mathrm{exposed}}$ for exposure, $G_{\mathrm{exposed}}-G_{\mathrm{selected}}$ for selection, and $G_{\mathrm{selected}}-G_{\mathrm{final}}$ for ranking.  These quantities place all three stages on the same target-normalized nDCG scale.

We additionally report the zero-exposure rate $N^{-1}\sum_i\mathbf{1}[E_i\cap Y_i=\varnothing]$ and the conditional retention ratios
\begin{equation}
\text{Gain kept}=\frac{G_{\mathrm{final}}}{G_{\mathrm{exposed}}},
\qquad
\text{Rank quality}=\frac{G_{\mathrm{final}}}{G_{\mathrm{selected}}},
\label{eq:funnel-ratios}
\end{equation}
for nonzero denominators.  Gain kept combines selection and ranking loss, while Rank quality isolates ordering conditional on the selected set; equivalently, Gain kept factors into $G_{\mathrm{selected}}/G_{\mathrm{exposed}}$ and Rank quality.  The zero-exposure rate identifies episodes in which no selector or ranker restricted to the observed candidates could recover a graded antecedent.  Together, these metrics localize where gain disappears along observed trajectories.  They do not by themselves identify whether an exposure loss was caused by the relevance hypothesis, query formulation, exploration policy, parametric memory, or search backend; such causal attributions require intervention.

%% file: sections/experiment.tex
\section{Experiments and Results}
\label{sec:experiments}


\subsection{Setup}
\label{sec:environment}

\textbf{Evaluation environment and information boundary}. Each \timemachine episode presents only $B_i$ through common \texttt{search\_papers} and \texttt{submit\_answer} schemas.  Every system receives the same brief, date filter, returned fields, search budget, output limit, and scorer.  The hidden target, antecedent tiers, and future citation contexts never enter the policy history, ensuring that no condition receives evaluator information during search.  The search backend, response schema, and episode budget are shared throughout.  We log each ordered query, returned identifier, tool error, and final submission to reconstruct first exposure and the complete funnel.  This common interface supports direct comparisons while leaving parametric memory, provider effects, and possible contamination as uncontrolled sources of variation.  Appendix~\ref{app:experiments} records the exact run manifest, model revisions, response settings, and call accounting.


\textbf{Evaluation models and metrics}.We evaluate Claude Opus 5 and Claude Sonnet 5~\cite{anthropic2026opus5,anthropic2026sonnet5}, GPT-5.6-Sol~\cite{openai2026gpt56}, Gemini 3.5 Flash and Gemini 3.1 Pro Preview~\cite{googledeepmind2026gemini35flash,googledeepmind2026gemini31pro}, and \sftmodel Base and its hindsight-SFT variant~\cite{qwenteam2026qwen36} on the automatically screened manifest with $D_i=t_{P_i}-3$ months.  SFT uses a disjoint earlier pool and a development set frozen before training.  Every condition follows the shared protocol above on the same ordered target manifest, and provider APIs only generate the next action.
The primary metric is failure-as-zero nDCG@10 \cite{jarvelin2002cumulated}, which rewards both high-gain recovery and placement near the top of the submitted ranking. We also report Recall@10, MRR, \anyhit, three- and five-hit rates, and total calls.  These measure the recovered label fraction, first-hit rank, any connection to the realized lineage, stronger portfolio recovery, and interaction use.  Multi-hit rates condition only on targets with enough labels, following the distinction between isolated relevance and diversified coverage \cite{carbonell1998mmr,clarke2008novelty}. Table~\ref{tab:benchmark-main} reports the aggregate point estimates.

\subsection{End-to-End and Portfolio Performance}
\label{sec:benchmark-results}
\label{sec:portfolio-results}

Table~\ref{tab:benchmark-main} reports the end-to-end comparison.  Claude Opus 5 leads with 0.284 nDCG@10, 0.215 Recall@10, and 0.723 \anyhit, followed by Claude Sonnet 5 at 0.189 nDCG@10; the ordering is broadly consistent across nDCG, recall, MRR, and \anyhit.  Yet even the leading system recovers only 21.5\% of labeled antecedents on average, indicating substantial headroom beyond identifying one plausible paper.

\input{tables/benchmark_results}

\begin{figure*}[!t]
  \centering
  \figureorplaceholder{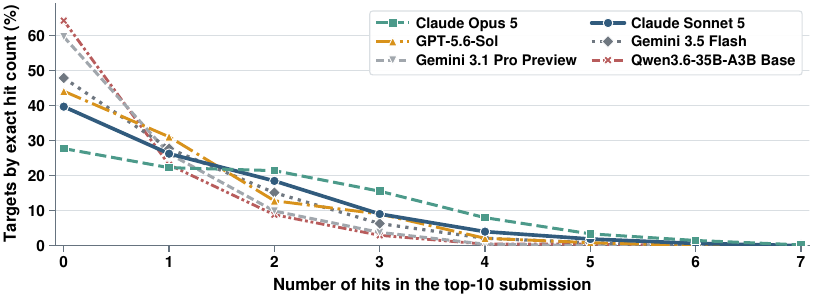}{0.98\textwidth}{2.15in}
  \caption{\textbf{Recovered antecedent counts across 476 targets.}  The horizontal axis is the number of graded antecedents recovered in the submitted top 10, and the vertical axis is the percentage of targets with exactly that count.  Failures and non-submissions enter the zero bin.  Concentration at zero and one shows that recovering an isolated antecedent is substantially more common than assembling a broader portfolio.}
  \label{fig:hit-distribution}
\end{figure*}

Binary success further obscures the difficulty of portfolio recovery.  Among eligible targets, the leading model's one-, three-, and five-hit rates decline from 72.3\% to 32.1\% and 6.7\%, and Figure~\ref{fig:hit-distribution} shows the same concentration at zero to two hits across systems.  Stronger agents therefore reduce complete misses more reliably than they assemble portfolios of complementary antecedents.  This pattern is not explained by interaction volume alone: Gemini 3.5 Flash averages 34.7 calls and obtains 0.151 nDCG@10, whereas Claude Opus 5 averages 21.1 calls and obtains 0.284.  What matters is both the evidence made available and how it is subsequently used, motivating the stage-wise analysis below.

\subsection{Stage-Wise Performance}
\label{sec:funnel-results}

Table~\ref{tab:funnel-main} reports the three nested quantities on a common normalized-gain scale.  Exposed gain measures the ranking potential made available by resource exposure, Selected gain measures how much potential remains after the agent decides what to submit, and final nDCG@10 additionally reflects the submitted order.  Their nesting, $G_{\mathrm{final}}\leq G_{\mathrm{selected}}\leq G_{\mathrm{exposed}}$, makes each successive gap attributable to one observed stage of the trajectory.

\input{tables/funnel_results}

\textbf{Resource exposure.}
Exposed gain ranges from 0.153 to 0.480, and the fraction of targets for which no graded antecedent is exposed ranges from 17.4\% to 56.1\%.  For every system, the largest absolute deficit relative to the normalized optimum occurs before selection: $1-G_{\mathrm{exposed}}$ is at least 0.520 and reaches 0.847.  Claude Opus 5 exposes the most gain and has the fewest complete misses, whereas \sftmodel Base exposes less than one third as much gain and misses every label on more than half of the targets.  Exposed gain and zero-exposure rate are complementary: the former captures the amount and grade of evidence made available across targets, whereas the latter isolates trajectories from which no labeled item could be recovered downstream.  Surfacing one low-gain antecedent can reduce the miss rate without producing a strong candidate portfolio.

\textbf{Selection.}
Selected gain ranges from 0.119 to 0.386.  The gap $G_{\mathrm{exposed}}-G_{\mathrm{selected}}$ measures useful evidence that was observed but omitted from the submission, independently of ordering.  Claude Sonnet 5 and GPT-5.6-Sol illustrate why this stage must be evaluated separately: they expose nearly identical gain (0.323 and 0.326), but Sonnet retains more of that potential (0.269 versus 0.233), contributing to its higher final nDCG@10 (0.189 versus 0.164).  Conversely, a small absolute selection gap is not by itself evidence of strong selection when little useful material was exposed upstream; stage quantities must be interpreted jointly.

\textbf{Ranking.}
Final nDCG@10 ranges from 0.082 to 0.284, while Rank quality ranges from 0.683 to 0.742.  The gap $G_{\mathrm{selected}}-G_{\mathrm{final}}$ isolates the loss induced by the submitted order, because Selected gain is the best score achievable by reordering the same submitted support.  Gemini 3.1 Pro Preview attains the highest Rank quality but exposes only 0.184 gain, demonstrating that effective ordering cannot compensate for evidence that was never surfaced.  A reranker can improve the use of an existing submission, but it cannot recover an antecedent absent from the observed trajectory.

\begin{figure*}[!t]
  \centering
  \figureorplaceholder{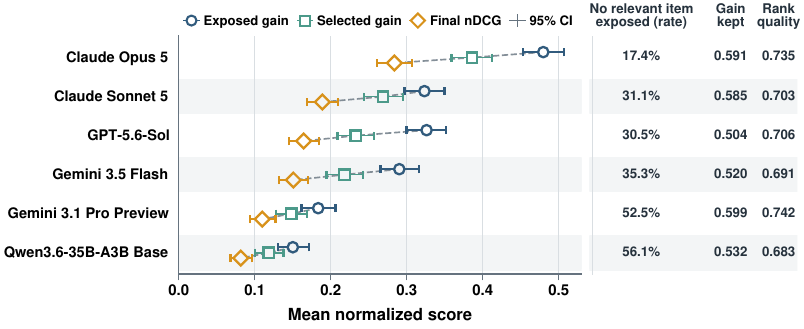}{0.98\textwidth}{2.65in}
  \caption{\textbf{Observed funnel across resource exposure, selection, and ranking.}  For each model, the three points show mean Exposed gain, Selected gain, and final nDCG@10; horizontal bars are 95\% target-bootstrap intervals.  The right columns report zero-exposure rate, Gain kept, and Rank quality.  Lower zero-exposure is better, whereas higher values are better for the other quantities.  Horizontal separation localizes the stage of observed losses; causal mechanisms require intervention.}
  \label{fig:funnel}
\end{figure*}

%% file: tables/benchmark_results.tex
\begin{table*}[!t]
  \caption{\textbf{Unified end-to-end comparison.}  All rows use the same targets, interface, search budget, failure policy, and scorer.  nDCG@10, Recall@10, MRR, and AnyHit@10 include failed or empty episodes as zero.  The multi-hit columns report exact recovery of at least one, three, or five graded antecedents among targets with enough labels; Total calls is the mean number of search and paper-access interactions.}
  \label{tab:benchmark-main}
  \centering
  \small
  \setlength{\tabcolsep}{4pt}
  \resizebox{\textwidth}{!}{%
  \begin{tabular}{lrrrrrrr}
    \toprule
    Model & nDCG@10 & Recall@10 & MRR & \anyhit
      & $\geq3$ hits
      & $\geq5$ hits & Total calls \\
    \midrule
    Claude Opus 5 & \textbf{0.284} & \textbf{0.215} & \textbf{0.476}
      & \textbf{0.723} & \textbf{0.321} & \textbf{0.067} & 21.1 \\
    Claude Sonnet 5 & 0.189 & 0.155 & 0.361
      & 0.603 & 0.175 & 0.033 & 17.7 \\
    GPT-5.6-Sol & 0.164 & 0.122 & 0.321
      & 0.559 & 0.134 & 0.011 & 26.3 \\
    Gemini 3.5 Flash & 0.151 & 0.120 & 0.299
      & 0.521 & 0.104 & 0.011 & 34.7 \\
    Gemini 3.1 Pro Preview & 0.110 & 0.082 & 0.258
      & 0.403 & 0.047 & 0.000 & 20.9 \\
    \sftmodel Base & 0.082 & 0.071 & 0.200
      & 0.357 & 0.042 & 0.006 & 21.0 \\
    \bottomrule
  \end{tabular}}
\end{table*}

%% file: tables/funnel_results.tex
\begin{table*}[!t]
  \caption{\textbf{Resource exposure, selection, and ranking funnel.}  Exposed gain oracle-sorts every graded antecedent encountered during search; Selected gain oracle-sorts the graded antecedents retained in the submission; final nDCG@10 additionally reflects their submitted order.  Gain kept and Rank quality are the retention ratios in Equation~\ref{eq:funnel-ratios}, while No relevant item exposed reports the zero-exposure rate.  Empty submissions remain zero, and stage gaps describe the observed trajectories.}
  \label{tab:funnel-main}
  \centering
  \small
  \renewcommand{\arraystretch}{1.08}
  \setlength{\tabcolsep}{4pt}
  \begin{tabular*}{\textwidth}{@{\extracolsep{\fill}}lrrrrrr@{}}
    \toprule
      & \multicolumn{2}{c}{Resource exposure}
      & Selection
      & Ranking
      & \multicolumn{2}{c}{Retention diagnostics} \\
    \cmidrule(lr){2-3}\cmidrule(lr){4-4}\cmidrule(lr){5-5}\cmidrule(l){6-7}
    Model & \shortstack{Exposed\\gain}
      & \shortstack{No relevant\\item exposed}
      & \shortstack{Selected\\gain}
      & \shortstack{Final\\nDCG@10}
      & \shortstack{Gain\\kept}
      & \shortstack{Rank\\quality} \\
    \midrule
    Claude Opus 5 & 0.480 & 17.4\%
      & 0.386 & 0.284 & 0.591 & 0.735 \\
    Claude Sonnet 5 & 0.323 & 31.1\%
      & 0.269 & 0.189 & 0.585 & 0.703 \\
    GPT-5.6-Sol & 0.326 & 30.5\%
      & 0.233 & 0.164 & 0.504 & 0.706 \\
    Gemini 3.5 Flash & 0.290 & 35.3\%
      & 0.218 & 0.151 & 0.520 & 0.691 \\
    Gemini 3.1 Pro Preview & 0.184 & 52.5\%
      & 0.148 & 0.110 & 0.599 & 0.742 \\
    \sftmodel Base & 0.153 & 56.1\%
      & 0.119 & 0.082 & 0.532 & 0.683 \\
    \bottomrule
  \end{tabular*}
\end{table*}

%% file: sections/results.tex
\input{tables/sft_results}

\subsection{Matched Hindsight SFT}
\label{sec:sft-protocol}

To test whether \inspire supports model development rather than merely separating fixed systems, we compare \sftmodel Base with a hindsight-SFT variant trained on a disjoint pool of earlier targets.  Hindsight SFT uses training-only outcome information to identify successful trajectories, replays their actions through the standard search interface, and retains only demonstrations whose submitted papers were exposed during replay.  The student is trained solely on the resulting deployment-legal transcripts: hidden relevance labels, teacher-side reasoning, and trajectory-selection scores are removed.  We then evaluate the base and SFT models on the same held-out targets with identical cutoffs, interfaces, search budgets, decoding settings, and scorers, making the adapter the only experimental change.  As shown in Table~\ref{tab:sft-main}, SFT raises exposed gain from 0.153 to 0.251 and reduces zero-exposure episodes from 56.1\% to 29.4\%, indicating more effective discovery of relevant resources.  It also raises selected gain from 0.119 to 0.219 and Rank quality from 0.683 to 0.759, showing that the model both retains more useful evidence and orders its final submission more effectively.  Together, these stage-wise improvements increase final nDCG@10 from 0.082 to 0.166.  The decomposition therefore reveals that hindsight supervision improves the full search pipeline rather than only its final output, demonstrating that \inspire provides actionable training signals as well as diagnostic evaluation; Appendix~\ref{sec:sft} details demonstration construction, leakage controls, training configuration, and paired uncertainty analysis.


%% file: tables/sft_results.tex
\begin{table}[!t]
  \caption{\textbf{Controlled benchmark-usability test.} The table reports changes in resource exposure, selection, and final ranking, allowing the end-to-end improvement to be localized, testing whether \inspire can produce useful supervision and measure held-out improvement.}
  \label{tab:sft-main}
  \centering
  \small
  \renewcommand{\arraystretch}{1.08}
  \setlength{\tabcolsep}{4pt}
  \begin{tabular*}{\textwidth}{@{\extracolsep{\fill}}lrrrrrr@{}}
    \toprule
      & \multicolumn{2}{c}{Search and gather}
      & \multicolumn{3}{c}{Filter and construct answer}
      & End-to-end \\
    \cmidrule(lr){2-3}\cmidrule(lr){4-6}\cmidrule(l){7-7}
    Condition & \shortstack{Exposed\\gain}
      & \shortstack{No relevant\\item exposed}
      & \shortstack{Selected\\gain}
      & \shortstack{Gain\\kept}
      & \shortstack{Rank\\quality}
      & \shortstack{Final\\nDCG@10} \\
    \midrule
    \sftmodel Base
      & 0.153 & 56.1\% & 0.119 & 0.532 & 0.683 & 0.082 \\
    \addlinespace[2pt]
    \sftmodel{} + hindsight SFT
      & \textbf{0.251} & \textbf{29.4\%} & \textbf{0.219}
      & \textbf{0.661} & \textbf{0.759} & \textbf{0.166} \\
    \bottomrule
  \end{tabular*}
\end{table}

%% file: sections/related_work.tex
\section{Related Work}
\label{sec:related}

\textbf{Scientific search.} Interactive IR models search as iterative clarification of a changing information need rather than one-shot lookup \cite{belkin1982ask,bates1989berrypicking,marchionini2006exploratory,saracevic1975relevance}, while literature-based discovery seeks non-obvious connections across separated bodies of work \cite{swanson1986fishoil}.  Language-model systems increasingly couple retrieval, browsing, reasoning, and tool use \cite{petroni2021kilt,lewis2020rag,nakano2021webgpt,yao2023react,schick2023toolformer,liu2024agentbench}; BrowseComp and DeepResearch Bench extend this trajectory to persistent clue-driven search and report-level evaluation \cite{wei2025browsecomp,du2025deepresearchbench}.  Scientific retrieval builds on probabilistic ranking and heterogeneous-domain evaluation \cite{robertson2009probabilistic,thakur2021beir}, as well as literature graphs, structured corpora, scientific encoders, and citation-aware representations or recommendation \cite{ammar2018semantic,lo2020s2orc,beltagy2019scibert,cohan2020specter,singh2023specter2,bhagavatula2018content}.  AutoResearchBench evaluates search against explicit multi-part conditions, whereas PaperPilot starts from an anchor and may elicit user intent \cite{autoresearchbench2026,li2026paperpilot}.  MIR and HyBIRD retrieve methodological inspirations from solution-redacted problems, and ResearchBench decomposes inspiration discovery and ranking within a curated candidate pool \cite{garikaparthi2025mir,yang2026hybird,liu2025researchbench}.  

\textbf{Hindsight supervision.} Freshness-aware retrieval motivates temporal isolation \cite{vu2023freshllms}.  PreScience searches a pre-target corpus while conditioning on future authors and their histories, SciPaths exposes a realized contribution, CUSP distinguishes plausible mechanisms from realized outcomes, and Future-Aligned Proposals supplies inspirations as input \cite{ajith2026prescience,chamoun2026scipaths,wu2026cusp,wang2026futurealigned}. Graded cumulative gain rewards relevance and rank \cite{jarvelin2002cumulated}, while diversified retrieval motivates evaluating portfolios rather than isolated hits \cite{carbonell1998mmr,clarke2008novelty}.  Studies of citation function and importance inform our tiered antecedent labels \cite{teufel2006automatic,valenzuela2015meaningful,macroberts1989problems,bornmann2008citation}.  Finally, learning with privileged information and hindsight reuse formalizes the use of train-only outcome signals \cite{vapnik2009new,andrychowicz2017hindsight}; our matched study uses these signals to select replay-valid demonstrations for standard supervised instruction tuning with LoRA \cite{ouyang2022training,hu2022lora}.

%% file: sections/limitations.tex
\section{Limitations}
\label{sec:limitations}

Extending \inspire along several dimensions merits future exploration.  Human missing-gold audits spanning uncited influences, older foundations, and alternative solution lineages could strengthen the current capped, recent arXiv citation labels \cite{macroberts1989problems,bornmann2008citation}.  Independent redaction studies, additional disciplines and search backends, and broader judge families could quantify residual semantic leakage, grading bias \cite{zheng2023judging}, and domain dependence.  Evaluations across students, training seeds, provider regimes, and controlled parametric-memory settings could test the generality of the SFT result and isolate the mechanisms behind observed funnel gaps.  Auditing ranked retrieval for cumulative-advantage effects \cite{merton1968matthew} would further clarify downstream impact.  \inspire is designed to evaluate search systems rather than researchers or intellectual provenance.

%% file: sections/conclusion.tex
\section{Conclusion}
\label{sec:conclusion}

Scientific search requires three coupled stages: \emph{resource exposure}, \emph{selection}, and \emph{ranking}.  \inspire evaluates them on the same logged trajectory through solution-redacted problems, date-gated open-corpus search, and hindsight-revealed graded antecedents.  Current agents recover isolated antecedents more readily than broader portfolios; their largest observed loss occurs during resource exposure, with further losses in selection and ranking.  Matched hindsight SFT improves all three stages under unchanged test-time information, showing that \inspire provides an actionable learning signal.  By distinguishing evidence that was never surfaced, exposed but omitted, or retained but poorly ranked, this decomposition enables targeted model development.

%% file: sections/ai_disclosure.tex
\subsection*{AI use statement}
\label{app:ai_use}

We used LLMs to support benchmark construction, including filtering candidate papers, generating solution-redacted problem briefs, grading cited antecedents, and auditing briefs for target identifiability. We also used an LLM teacher to generate synthetic search demonstrations for supervised fine-tuning and evaluated LLM-based search agents, as detailed in the methodology and appendices. For manuscript preparation, we used generative AI tools for language polishing and writing assistance. We reviewed and revised the AI-assisted text and take full responsibility for the final paper, including its claims, results, and accompanying artifacts.

%% file: sections/dataset_details.tex
\section{Dataset Construction}
\label{app:data}

\subsection{Overview}

Overall, \inspire comprises recent influential computer-science target papers resolved to arXiv. Each retained instance contains a solution-redacted problem brief, a target-specific public-record cutoff, and automatically graded pre-cutoff cited antecedents with tier and role annotations. Table~\ref{tab:dataset} reports the train and evaluation statistics separately.

\input{tables/dataset_statistics}

A release manifest records the ordered train, development, evaluation, and diagnostic lists; SHA-256 hashes of instance files; construction settings; judge model identifiers; creation time; and the retrieval-response manifest.

\subsection{Target Discovery}

The builder issues Semantic Scholar bulk searches across computer-science and machine-learning subfields, deduplicates arXiv identifiers across topic queries, and uses arXiv metadata as the authority for title, abstract, category, and v1 date. Within each topic it ranks a candidate using
\begin{equation}
  h(P)=\text{influential-citation-count}(P)
       +\frac{\text{citation-count}(P)}{100}.
\end{equation}
This heuristic orders target discovery. A model-based filter removes surveys and reviews, after which the builder rejects papers without parseable arXiv HTML or at least two graded pre-cutoff antecedents. These choices define a population of recent, visible, sufficiently cited work.

\subsection{Reference Grading}

The target's HTML bibliography supplies candidate arXiv identifiers, which are normalized and batch-resolved. A reference with v1 date $t_r\ge D$ is stored as an excluded concurrent reference and never enters the label set. Tier judges receive the target abstract, candidate metadata, and target citation context when available. Their role-sensitive grading is motivated by prior evidence that citation contexts encode heterogeneous functions and levels of importance \cite{teufel2006automatic,valenzuela2015meaningful}. For every candidate, the instance retains successful votes, consensus tier and role, rationale, and agreement count. Failed voters do not abort construction, so some records have fewer than three valid votes; the release reports those denominators. The active discriminativeness filter retains references within four years of the target. After T0 removal, references are sorted by configured gain and capped at ten. Older foundations and overflow remain retrievable but unlabeled. Predictions outside $Y$ receive zero benchmark credit, though they may remain scientifically useful. The declared conceptual gains for T1/T2/T3 are $1/3/7$; infrastructural gains are $0.5/2/4$. Sensitivity analyses cover binary relevance, T2/T3-only scoring, and alternate monotone maps.

\subsection{Brief Generation and Screening}

The generator receives the target title and at most 1,800 characters of its abstract. It produces a detailed and a terse brief while being instructed to remove the target name, mechanism, and target-specific results. In the current evaluation snapshot, the detailed variant averages 179.5 whitespace-delimited words and the terse variant 103.2. A no-tool auditor sees the detailed brief and returns a target guess, confidence, and rationale. A positive guess with confidence at least 0.5 flags the instance; a separate lexical check records target-title token overlap but does not determine the split. The audit defines reproducible \emph{automatic brief-clean} and flagged partitions; main results use the first and report all-target and flagged views alongside it. Because the audit is model-based, these slices characterize screening outcomes rather than independently validating every redaction \cite{zheng2023judging}.

\begin{figure*}[!t]
  \centering
  \figureorplaceholder{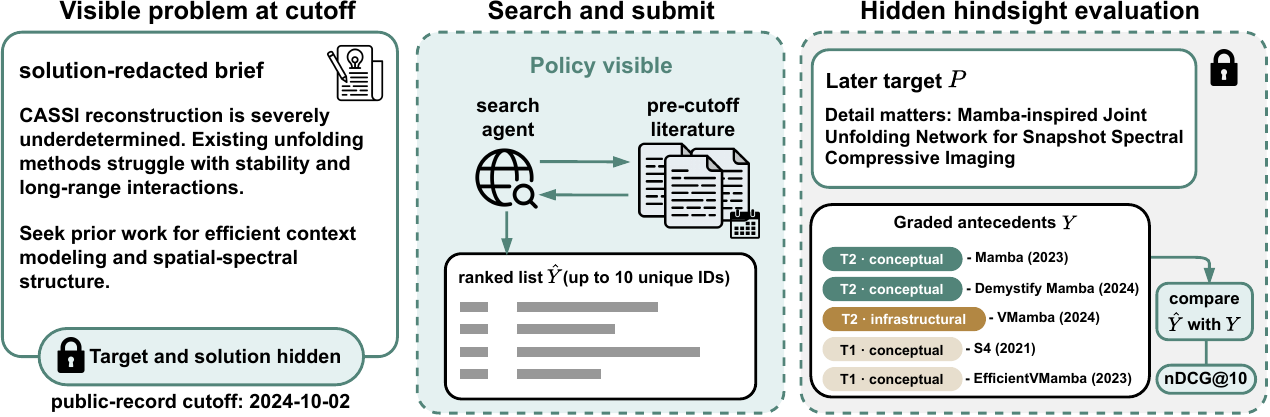}{0.97\textwidth}{1.75in}
  \caption{\textbf{Development example and automatic audit outcome.}  The agent receives a shortened CASSI brief and public-record cutoff; the later target and colored antecedent labels are evaluator-only.  The target-identity audit did not name the paper but inferred a likely Mamba/state-space mechanism, illustrating the residual mechanism-level leakage captured by the audit.  Tiers and roles come from \texttt{arxiv:2501.01262}.}
  \label{fig:task-example}
\end{figure*}

Figure~\ref{fig:task-example} presents a development example retained from benchmark construction.  The target-identity auditor did not name the paper, while its rationale inferred a likely Mamba or state-space mechanism.  The example illustrates the task interface, hindsight scoring, and the residual mechanism-level clues captured by the automatic audit.

%% file: tables/dataset_statistics.tex
\begin{table*}[ht]
  \caption{\textbf{\inspire dataset statistics.}  Train and evaluation partitions are reported separately.  Antecedents are resolved pre-cutoff references cited by the later targets and automatically graded as T3 (most central), T2, or T1, operationalizing each target's realized citation lineage.  The final columns summarize brief length and target-release years.}
  \label{tab:dataset}
  \centering
  \small
  \setlength{\tabcolsep}{4.5pt}
  \resizebox{\textwidth}{!}{%
  \begin{tabular}{lrrrrrrrr}
    \toprule
    Split & Targets & Graded antecedents & Median/target & T3 & T2 & T1 & Mean brief words & Target years \\
    \midrule
    Train & 1{,}549 & 10{,}268 & 7 & 430 & 2{,}905 & 6{,}933 & 183.4 & 2023 to 2024 \\
    Evaluation & 476 & 3{,}574 & 10 & 179 & 1{,}244 & 2{,}151 & 179.3 & 2025 to 2026 \\
    \bottomrule
  \end{tabular}}
\end{table*}

%% file: sections/experiment_details.tex
\section{Experimental Details}
\label{app:experiments}

\subsection{Shared Protocol}

The run manifest contains the ordered target lists and instance hashes; target-specific cutoffs; search-backend and response-schema identifiers; system prompt, user prompt, tool schema, and scorer hashes; the fixed search budget; model revisions and decoding settings; software commit; and run creation time.  The same interface implementation and budget are used for all five closed-source models, \sftmodel Base, and \sftmodel with SFT.  A development replay checks date filtering, identifier normalization, output limits, and failure-as-zero scoring before a row is admitted to the main table. Every episode record stores model, target, protocol hash, ordered queries, ordered returned IDs, first exposure of each unique ID, tool errors, final ranking, terminal status, calls, tokens, latency, and cost where available.  A comparability check verifies equality of target manifest, brief text, returned fields, cutoff logic, search budget, submission schema, and scorer across rows. The reporter emits expected, attempted, completed, submitted, and scored counts before computing any aggregate.

\subsection{Model Evaluation}

Each recorded checkpoint is evaluated once on every ordered evaluation target. The primary automatic-clean view is joined by target identifier before aggregation; the all-target and automatic-flagged views reuse the same episodes. Outputs are canonicalized to arXiv base IDs, duplicate submissions keep their first rank, and empty or invalid submissions remain zero.  The main table contains no model whose policy-visible interface or search budget differs from the others. The model roster is Claude Opus 5, Claude Sonnet 5, GPT-5.6-Sol, Gemini 3.5 Flash, Gemini 3.1 Pro Preview, \sftmodel Base, and \sftmodel with hindsight-distilled SFT.  Exact provider snapshot identifiers are part of the artifact manifest.  If a provider revision cannot be pinned, its request time and returned revision metadata are retained and the reproducibility limitation is stated rather than replacing the row post hoc.

The target is the statistical unit. End-to-end comparisons use paired target-bootstrap intervals over the shared manifest, following paired resampling practice for system-level comparisons \cite{koehn2004statistical}. Automatic slices by split, tier, role, topic, label count, and training-antecedent overlap are descriptive and always report their denominator.

\subsection{Trajectory Metrics}

The trajectory reporter reconstructs the first 50 and unrestricted unique exposure sets, final submitted ranking, and first exposure turn for every ID. It reports relevant-item recall among the first 50 candidates, unrestricted exposure recall, zero-exposure targets, relevant-item retention conditional on positive exposure, exposed and selected gain, gain kept, rank quality, duplicate queries, malformed calls, stopping behavior, and unique IDs per call. Unit checks reconcile the final-stage mean with the main nDCG table and verify that selected gain lies between zero and exposed gain, and exposed gain lies below one, whenever only exposed IDs can be submitted. A cross-target join asks whether a target-specific missed antecedent appears in another trajectory produced through the same search backend.  This analysis measures backend reachability; target-specific association and selection remain untested by the join.

\subsection{SFT Configuration}

The SFT student is \sftmodel.  For automatic-clean training instances, the teacher receives the visible brief plus a private block containing up to 12 highest-tier antecedent titles and truncated abstracts, without IDs.  It generates four candidate trajectories.  Each candidate is replayed through the ordinary deployment prompt and search interface, then scored and scanned for private-string copying, paper-identifying queries, post-cutoff results, malformed actions, and unsurfaced submissions.  At most one highest-scoring replay-valid trajectory above a development-frozen threshold is retained per target.

The final corpus manifest records source instance, teacher revision, random seed, score, action count, rejection flags, transcript hash, and selected status.  The SFT configuration uses action-token loss, three epochs, LoRA rank 16, scale 32, zero dropout, and learning rate $5\times10^{-6}$; LoRA supplies the parameter-efficient adapter design \cite{hu2022lora}.  The base-checkpoint hash, tokenizer and chat template, target modules, batching, optimizer, schedule, precision, maximum sequence length, and software/hardware environment are also recorded.  Base and SFT evaluation differ only by adapter. 

%% file: sections/hindsight_sft.tex
\section{Hindsight SFT Details}
\label{sec:sft}

This appendix provides the demonstration-construction, information-boundary, and uncertainty details supporting the matched study in Section~\ref{sec:sft-protocol}.

\paragraph{Training protocol.}
The student and matched base model are \textbf{\sftmodel}, and supervised fine-tuning is the only training stage.  For targets in the disjoint training pool that pass the same automatic screen, a frozen GPT-5.5 teacher~\cite{openai2026gpt55} may inspect training-only antecedent tiers, titles, and abstracts, but never their identifiers.  Its proposed actions are replayed through the ordinary tools, and an identifier may be submitted only after the replay exposes it.  The highest-scoring replay-valid trajectory above a development-frozen threshold becomes the demonstration; automatic scans reject target or antecedent copying, post-cutoff results, and unsurfaced submissions.  The student receives only the deployment transcript---the redacted brief, prior actions, and ordinary search results---while private teacher text and demonstration-selection scores are removed.  Privileged hindsight therefore selects successful training behavior without expanding the student's test-time observations or actions, following the train-only information boundary of privileged learning and hindsight replay \cite{vapnik2009new,andrychowicz2017hindsight}.  Training uses standard LoRA SFT \cite{hu2022lora}; Appendix~\ref{app:experiments} records the corpus, checkpoint, and optimization configuration.

\begin{figure*}[!t]
  \centering
  \figureorplaceholder{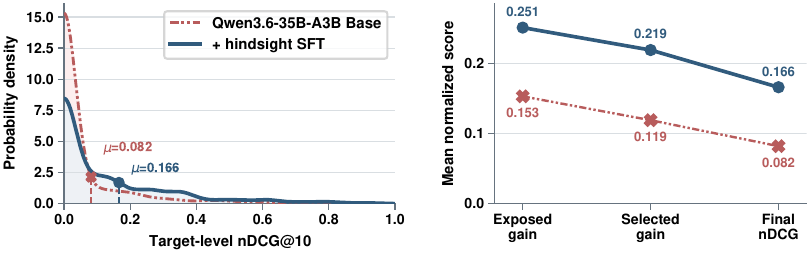}{0.98\textwidth}{2.15in}
  \caption{\textbf{Held-out response to benchmark-derived SFT.}  The left panel compares boundary-reflected Gaussian kernel densities of matched Base and SFT failure-as-zero nDCG@10 values, including the probability mass at zero.  The right panel traces aggregate Exposed gain, Selected gain, and final nDCG@10 under the same held-out protocol.  Improvements at all three points indicate that the controlled intervention changes both candidate discovery and use of the discovered evidence.}
  \label{fig:sft-comparison}
\end{figure*}

\paragraph{Matched evaluation.}
At evaluation, \sftmodel Base and the SFT model use the same held-out targets, interface, cutoff, search budget, decoding, failure policy, and scorer; the adapter is the only changed factor.  Table~\ref{tab:sft-main} reports the stage-wise and end-to-end point estimates, while Figure~\ref{fig:sft-comparison} shows the paired target-level score distributions.  Final nDCG@10 increases from 0.082 to 0.166, a paired difference of 0.084 with a 95\% paired target-bootstrap interval of $[0.055,\,0.115]$.  This interval quantifies variation over the shared target set; training-seed variation remains outside the reported uncertainty when only one frozen adapter is available.

%% file: sections/oracle_search_progress.tex
\section{Exposure Gain over Search}
\label{app:oracle-progress}

The aggregate exposure funnel measures how much graded gain is available at the end of a trajectory, but it does not show when that gain enters the candidate set. We therefore compute a prefix-based oracle diagnostic. For condition $c$, let $E_i^{(c)}(p)$ contain the unique papers exposed by target $i$ up to normalized search progress $p\in[0,1]$, where $E_i^{(c)}(0)=\varnothing$ and $E_i^{(c)}(1)=E_i^{(c)}$. Using the target-level oracle in Equation~\ref{eq:ranking-potential}, the progress curve is
\begin{equation}
  G_{\mathrm{exposed}}^{(c)}(p)
  =\frac{1}{N_c}\sum_{i=1}^{N_c}\phi_i\!\left(E_i^{(c)}(p)\right).
  \label{eq:oracle-progress}
\end{equation}
At each checkpoint, this quantity asks how well the graded antecedents exposed so far could be ranked if their gains were known and all ungraded papers were discarded. The curves are consequently nondecreasing by construction, and their endpoints reproduce the Exposed gain values in Tables~\ref{tab:funnel-main} and~\ref{tab:sft-main}. Normalizing progress within each trajectory makes timing profiles comparable despite different trajectory lengths, while absolute calls and returned-paper counts can still differ across systems. The oracle retrospectively describes the value accumulated along each observed path; budget interventions require separate experiments.

\begin{figure}[t]
  \centering
  \includegraphics[width=\textwidth]{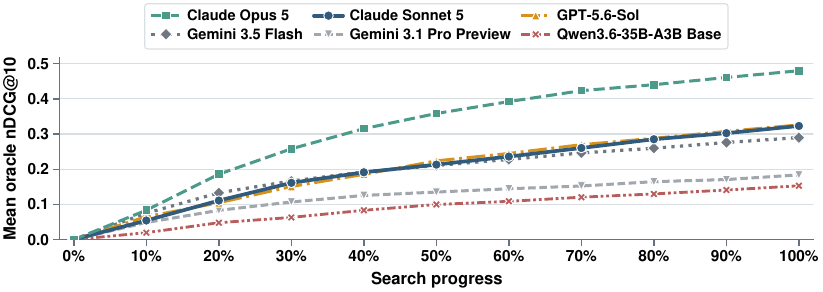}
  \caption{\textbf{Oracle nDCG@10 accumulated over normalized search progress.} At each checkpoint, graded antecedents in the observed exposure prefix are ordered by their hindsight gains. The endpoint of each curve is therefore the best nDCG@10 supported by the complete exposed set, before losses from final selection and ordering.}
  \label{fig:oracle-progress-models}
\end{figure}

Figure~\ref{fig:oracle-progress-models} reveals substantial differences not only in final exposure, but also in the rate at which useful evidence becomes available. Claude Opus 5 separates early, reaching approximately 0.18 oracle nDCG@10 after 20\% of its normalized search and 0.36 at the halfway point before ending at 0.480. Its advantage therefore reflects a sustained accumulation of graded antecedents rather than a single late discovery. Claude Sonnet 5 and GPT-5.6-Sol, by contrast, follow nearly overlapping trajectories through much of the latter half of search and finish at 0.323 and 0.326. Their similar exposure potential, together with Sonnet's higher final nDCG@10 of 0.189 versus 0.164, strengthens the funnel result that retention and ordering can separate systems even when search supplies comparable candidate sets.

No curve reaches an early plateau. At 50\% normalized progress, the systems have accumulated approximately 64--74\% of the oracle gain available at the end of their own trajectories, leaving roughly one quarter to one third of eventual exposed gain to arrive during the second half. Later search is therefore not uniformly redundant, even for the strongest model. At the same time, the low curves for Gemini 3.1 Pro Preview and \sftmodel Base remain low throughout search, despite continuing to increase. This pattern indicates that their exposure deficit is not confined to premature stopping: relative to the stronger systems, useful antecedents enter the candidate set at a lower rate across most of the observed trajectory. The curves localize this deficit along realized behavior; experiments targeting query choice, exploration strategy, parametric knowledge, and the retrieval backend can isolate its mechanism.

\begin{figure}[t]
  \centering
  \includegraphics[width=\textwidth]{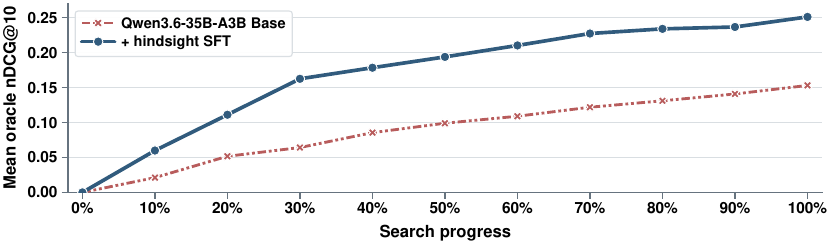}
  \caption{\textbf{Search-progress profile for the matched Base and hindsight-SFT conditions.} The oracle is recomputed on each condition's cumulative exposure prefix under the shared held-out protocol. Curve separation measures when the adapter makes additional graded gain available, independently of subsequent selection and submitted ordering.}
  \label{fig:oracle-progress-sft}
\end{figure}

Figure~\ref{fig:oracle-progress-sft} shows that the hindsight-SFT condition dominates the Base exposure curve at every nonzero checkpoint. The separation emerges early: by 30\% progress, SFT has accumulated approximately 0.16 oracle nDCG@10, already slightly more than the Base model obtains from its complete trajectory (0.153). The absolute gap is about 0.10 at this checkpoint and remains of similar magnitude thereafter, ending at 0.251 versus 0.153. Thus, the exposure improvement reported in Table~\ref{tab:sft-main} is not explained only by a small number of discoveries near the end of search. Instead, the observed SFT policy reaches higher-value candidate sets earlier and preserves that advantage throughout the trajectory.

Both conditions nevertheless continue to acquire useful evidence after the midpoint. From 50\% to 100\% progress, oracle nDCG@10 increases by approximately 0.06 for both Base and SFT. Hindsight SFT therefore shifts the entire availability profile upward without exhausting the value of continued exploration. The remaining difference between the SFT endpoint oracle of 0.251 and its final nDCG@10 of 0.166 also confirms that improved exposure leaves downstream headroom: some available gain is still lost during selection and ordering. Taken together with the matched funnel analysis, the progress curves show that the adapter changes observed search behavior early enough to expose better antecedents while also improving their eventual use. Identifying the learned query or appraisal decisions responsible for this shift requires a targeted intervention.

%% file: sections/llm_prompt_templates.tex
\section{Prompt Templates}
\label{app:prompts}

This section reports the natural-language templates used by the benchmark
implementation.  Text in braces denotes a value inserted at runtime.  System
and user messages are identified explicitly; tool results and the function
schemas supplied alongside the prompts are not natural-language templates and
are therefore not reproduced here.


\noindent\textbf{Full-tool benchmark agent. }
The runner uses this system message when the agent can search, fetch a paper's
full text, and submit a ranked answer.

\begin{leftbar}
\begin{Verbatim}[
fontsize=\small,
breaklines=true,
breaksymbolleft={},
breaksymbolright={}
]
You are a research assistant operating inside a point-in-time literature search environment frozen to a cutoff date. You are given a research brief describing an open problem as it stood before any solution existed. Your job is to find the prior papers most worth reading to make progress on that problem.

Rules:
- Use ONLY the provided tools. You have no web access.
- Search with varied keyword queries, read abstracts and full text as needed.
- Then call submit_answer with a ranked list of paper ids (most load-bearing first), each with a one-line rationale. Aim for 5-10 papers.
- Prioritize papers that supply the core mechanism or concept the solution would build on, over merely topical or background papers.
- You MUST finish by calling submit_answer.
\end{Verbatim}
\end{leftbar}


\noindent\textbf{Solution-redacted brief generation. }
The pinned Claude Opus 4.8 construction judge receives the target metadata and
generates the detailed and terse brief variants.

\begin{leftbar}
\begin{Verbatim}[
fontsize=\small,
breaklines=true,
breaksymbolleft={},
breaksymbolright={}
]
[System]
You write research briefs for the INSPIRE benchmark. A brief states an open research problem AS IT STOOD BEFORE any solution existed -- the motivation, the limitations of prior approaches, and the desired properties of a solution. It must NOT reveal the solution. Redact: the target paper's method name, the specific solution mechanism, and any result/benchmark/dataset names unique to the target. Write in the first person plural ('we want...'). End with: 'What prior work should we read to make progress on this problem?'

[User]
TARGET PAPER
 title: {target_title}
 abstract: {target_abstract}

\end{Verbatim}
\end{leftbar}

\noindent\textbf{Prior-reference tier and role judging. }
Claude Opus 4.6, GPT-5.5, and Gemini 3.1 Pro Preview each receive the same
system and user templates.  Candidate abstracts are truncated to 600
characters, the target abstract to 1,500 characters, and the target full text
is included only when available.  The global-citation field is likewise
conditional.

\begin{leftbar}
\begin{Verbatim}[
fontsize=\small,
breaklines=true,
breaksymbolleft={},
breaksymbolright={}
]
[System]
You are an expert research-literature judge for the INSPIRE benchmark. Given a TARGET paper's abstract and a candidate PRIOR reference, you decide how load-bearing that reference was for the target's core contribution.

Tier scale (how essential to the target's central mechanism/idea):
  T3 = backbone: the target directly builds its core method on this work.
  T2 = component: supplies a key concept/tool the method adapts or compares against.
  T1 = background: foundational or comparison context, not the mechanism.
  T0 = incidental: tangential or boilerplate citation, non-relevant.
Type:
  conceptual = supplies an idea/mechanism/insight the target reasons with.
  infrastructural = supplies tooling/architecture/data/benchmark the target uses.

CRITICAL -- reward SPECIFIC, NON-OBVIOUS antecedents over ubiquitous foundations. A famous, very-highly-cited base paper (a canonical architecture, a seminal method everyone in the field cites) is a SAFE, low-value seed: any keyword search trivially surfaces it, so it does not test retrieval skill. Reserve T3/T2 for the close, specialized predecessors a strong researcher would find and a weak one would miss; push generic foundations toward T1/T0 even if technically cited. When the TARGET's full text is provided, ground each decision in HOW the target actually uses the reference in-context (the section and sentences that cite it) rather than guessing from titles alone; a reference whose method/result the target leans on in its core sections outranks one mentioned only in passing.

[User]
TARGET PAPER
 title: {target_title}
 abstract: {target_abstract[:1500]}

[Included only when available:]
TARGET FULL TEXT (use to see how each reference is cited in-context):
{target_fulltext}

CANDIDATE PRIOR REFERENCES (cited by the target):
[0] id={reference_0.arxiv_id} ({reference_0.date})  [global_citations={reference_0.cited_by_count}]
    title: {reference_0.title}
    abstract: {reference_0.abstract[:600]}
...

For EACH reference, output its tier and type. Respond with ONLY a JSON array, one object per reference, in the same order, each as:
{"index": <int>, "arxiv_id": "<id>", "tier": "T0|T1|T2|T3", "type": "conceptual|infrastructural", "rationale": "<one sentence>"}
\end{Verbatim}
\end{leftbar}


\noindent\textbf{Privileged trajectory teacher. }
The GPT-5.5 teacher system message begins with below verbatim and
appends the instructions and optional private ingredient block below.  In the
reported privileged setting, up to 12 references are sorted by tier and
discriminativeness; each abstract is whitespace-normalized and truncated to
1,200 characters.

\begin{leftbar}
\begin{Verbatim}[
fontsize=\small,
breaklines=true,
breaksymbolleft={},
breaksymbolright={}
]
{SEARCH_ONLY_SYSTEM_PROMPT}

You are writing a high-quality tool-use demonstration for a smaller student model. Make exactly ONE function call per response. Use the research brief to devise effective searches. Operate through search_papers and submit only IDs actually returned by search.

Optional private candidate ingredients are provided below. Use them only to improve search strategy; never copy or submit an ID unless search_papers returned it.

PRIVATE CANDIDATE INGREDIENTS (teacher-only; not part of the student prompt):
1. [{seed_1.tier}/{seed_1.type}] {seed_1.title}
Abstract: {seed_1.abstract[:1200]}

...

[User]
RESEARCH BRIEF:
{brief_text}

Find and submit your ranked reading list.
\end{Verbatim}
\end{leftbar}